\documentclass[12pt] {article}
\usepackage{psfrag}
\usepackage{graphicx}
\usepackage{latexsym,amsfonts}
\usepackage{amssymb}
\begin{document}
\setcounter{page}{1}
\def\theequation{\arabic{section}.\arabic{equation}}
\def\theequation{\thesection.\arabic{equation}}
\setcounter{section}{0}

\title{Massless Thirring fermion fields in the boson field
representation}

\author{M. Faber\thanks{E--mail: faber@kph.tuwien.ac.at, Tel.:
+43--1--58801--14261, Fax: +43--1--5864203} ~~and~
A. N. Ivanov\thanks{E--mail: ivanov@kph.tuwien.ac.at, Tel.:
+43--1--58801--14261, Fax: +43--1--5864203}~\thanks{Permanent Address:
State Technical University, Department of Nuclear Physics, 195251
St. Petersburg, Russian Federation}}

\date{\today}

\maketitle
\vspace{-0.5in}
\begin{center}
{\it Atominstitut der \"Osterreichischen Universit\"aten,
Arbeitsbereich Kernphysik und Nukleare Astrophysik, Technische
Universit\"at Wien, \\ Wiedner Hauptstr. 8-10, A-1040 Wien,
\"Osterreich }
\end{center}

\begin{center}
\begin{abstract}
We show that the boson field representation of the massless fermion
fields, suggested by Morchio, Pierotti and Strocchi in
J. Math. Phys. {\bf 33}, 777 (1992) for the operator solution of the
massless Thirring model, agrees completely with the existence of the
chirally broken phase in the massless Thirring model revealed in EPJC
{\bf 20}, 723 (2001) and hep--th/0112183, when the free massless boson
fields are described by the quantum field theory,  free of infrared
divergences in 1+1--dimensional space--time, formulated in
hep--th/0112184 and hep--th/0204237.

\end{abstract}
\end{center}

\newpage

\section{Introduction}
\setcounter{equation}{0}

\hspace{0.2in} In our recent publications [1--4] we have revealed a
non--perturbative phase of spontaneously broken continuous symmetry in
the massless Thirring model [1,4] and the free massless (pseudo)scalar
field theory defined in 1+1--dimensional space--time. We have shown
that the massless Thirring model [5], invariant under chiral
$U(1)\times U(1)$ symmetry, is unstable under chiral symmetry breaking
[1,4]. We have proved that the spontaneously broken phase of the
massless Thirring model is fully characterized by a non--vanishing
fermion condensate and the wave function of the ground state of the
chirally broken phase coincides with the wave function of the ground
state of the superconducting phase of the BCS--theory of
superconductivity [1]. In the quantum field theory of a free massless
(pseudo)scalar field the symmetry broken phase is characterized by a
non--vanishing spontaneous magnetization and the wave function of the
ground state is infinitely degenerate [2,3]. Goldstone bosons are the
quanta of the free massless (pseudo)scalar field [1--3,6].

We would like to emphasize that these results cannot be considered as
counterexamples to the well--known Mermin--Wagner--Hohenberg theorem
[7] asserting the vanishing of {\it long range order} in quantum field
theories in two dimensions. Indeed, as has been pointed out by the
authors [7], the absence of the {\it long--range order} can be
inferred only for non--zero temperature $T\neq 0$ [7] and no
conclusion about its value can be derived for $T=0$ [7]. Unlike the
quantum field theories treated by Mermin, Wagner and Hohenberg [7],
the massless Thirring model [1,4] and the quantum field theory of a
free massless (pseudo)scalar field [1--3] are formulated at zero
temperature $T = 0$ and fermion {\it condensation} and spontaneous
{\it magnetization} have been found at $T=0$. Hence, these results go
beyond the scope of the applicability of the Mermin--Wagner--Hohenberg
theorem [7].

Coleman's theorem, asserting the absence of Goldstone bosons and a
spontaneously broken continuous symmetry by example of the quantum
field theory of a free massless (pseudo)scalar field [8], has been
recently critically discussed in our paper [3]. We have shown that the
fulfillment of this theorem would lead to the vanishing not only
vacuum expectation value of the variation of the free massless
(pseudo)scalar field, required by the Goldstone theorem [9] in the
case of the absence of a spontaneously broken continuous symmetry, but
of all Wightman functions [3].

The problem of infrared divergences is one of the main problems in
the quantum field theory of a free massless (pseudo)scalar field
[10--12]. The infrared divergences appear in the two--point Wightman
functions
\begin{eqnarray}\label{label1.1}
\hspace{-0.3in}D^{(+)}(x) &=& \langle
0|\vartheta(x)\vartheta(0)|0\rangle =
\frac{1}{2\pi}\int^{\infty}_{-\infty}\frac{dk^1}{2k^0}\,e^{\textstyle
-\,i\,k\cdot x} = - \frac{1}{4\pi}\,{\ell n}[-\mu^2x^2 +
i\,0\cdot\varepsilon(x^0)], \nonumber\\ \hspace{-0.3in}D^{(-)}(x) &=&
\langle 0|\vartheta(0)\vartheta(x)|0\rangle =
\frac{1}{2\pi}\int^{\infty}_{-\infty}\frac{dk^1}{2k^0}\,e^{\textstyle
+\,i\,k\cdot x} = - \frac{1}{4\pi}\,{\ell n}[-\mu^2x^2 -
i\,0\cdot\varepsilon(x^0)],
\end{eqnarray}
where $\varepsilon(x^0)$ is the sign function, $x^2 = (x^0)^2 -
(x^1)^2$, $k\cdot x = k^0x^0 - k^1x^1$, $k^0 = |k^1|$ is the energy of
a free massless (pseudo)scalar quantum with momentum $k^1$ and $\mu$
is the infrared cut--off reflecting the infrared divergence of the
Wightman function (\ref{label1.1})\,\footnote{Further in order to
underscore the dependence of the two--point Wightman functions on the
scale $\mu$ we will denote $D^{(\pm)}(x) \to D^{(\pm)}(x; \mu)$.}. For
the calculation of the two--point Wightman functions (\ref{label1.1})
one should use the expansion of the massless (pseudo)scalar field
$\vartheta(x)$ into plane waves [1--3]
\begin{eqnarray}\label{label1.2}
\vartheta(x) =
\int^{\infty}_{-\infty}\frac{dk^1}{2\pi}\,\frac{1}{2k^0}\,
\Big(a(k^1)\,e^{\textstyle -i\,k\cdot x} +
a^{\dagger}(k^1)\,e^{\textstyle i\,k\cdot x}\Big),
\end{eqnarray}
where $a(k^1)$ and $a^{\dagger}(k^1)$ are annihilation and creation
operators obeying the standard commutation relation
\begin{eqnarray}\label{label1.3}
[a(k^1), a^{\dagger}(q^1)] = (2\pi)\,2k^0\,\delta(k^1 - q^1).
\end{eqnarray}
Klaiber was the first [10] who asserted the importance of the problem
of infrared divergences of the two--point Wightman (\ref{label1.1})
for the solution of the massless Thirring model: {\it If one wants to
solve the Thirring model, one has to overcome this problem.} 

In Ref.[2] we have suggested a recipe for the struggle against the
infrared divergences of the two--point Wightman functions
(\ref{label1.1}). This recipe concerns the constraint on the external
source $J(x)$ of a free massless (pseudo)scalar field $\vartheta(x)$
in the path integral formulation of the quantum field theory of a free
massless (pseudo)scalar field. Let $Z[J]$, the generating functional
of Green functions of a free massless (pseudo)scalar field
$\vartheta(x)$, be defined by
\begin{eqnarray}\label{label1.4}
\hspace{-0.15in}Z[J] = \Big\langle 0\Big|{\rm T}\Big(e^{\textstyle
i\int d^2x\,\vartheta(x)J(x)}\Big)\Big|0\Big\rangle = \int {\cal
D}\vartheta\,e^{\textstyle i\int d^2x\,[{\cal L}(x) +
\vartheta(x)J(x)]},
\end{eqnarray}
where ${\rm T}$ is a time--ordering operator.  The Lagrangian of the
$\vartheta$--field ${\cal L}(x)$ reads
\begin{eqnarray}\label{label1.5}
{\cal L}(x) =
\frac{1}{2}\,\partial_{\mu}\vartheta(x)\partial^{\mu}\vartheta(x).
\end{eqnarray}
It is invariant under field translations [1--3,6]
\begin{eqnarray}\label{label1.6}
 \vartheta(x) \to \vartheta\,'(x) = \vartheta(x) + \alpha,
\end{eqnarray}
where $\alpha$ is an arbitrary parameter $\alpha \in \mathbb{R}^1$
[1--3]\,\footnote{As has been shown in [1] the parameter $\alpha = -
2\alpha_{\rm A}$ is related to the chiral phase $\alpha_{\rm A}$ of
global chiral rotations of Thirring fermion fields.}. Any correlation
function of the free massless $\vartheta$--field can be calculated in
terms of functional derivatives of $Z[J]$ with respect to the external
source $J(x)$ [2]
\begin{eqnarray}\label{label1.7}
&&G(x_1,\ldots,x_n;y_1,\ldots,y_p) = \langle
0|F(\vartheta(x_1),\ldots,\vartheta(x_n);\vartheta(y_1),\ldots,
\vartheta(y_p))|0\rangle = \nonumber\\ &&=
F\Big(-i\frac{\delta}{\delta J(x_1)},\ldots,-i\frac{\delta}{\delta
J(x_n)};- i\frac{\delta}{\delta J(y_1)},\ldots,-i\frac{\delta}{\delta
J(y_p)}\Big)Z[J]\Big|_{\textstyle J = 0}.
\end{eqnarray}
One encounters the problem of the calculation of these correlation
functions in connection with the calculation of correlations functions
in the massless Thirring model [1,2], since the bosonized version of
the massless Thirring model reduces to the quantum field theory of the
free massless pseudoscalar field $\vartheta(x)$ described by the
Lagrangian (\ref{label1.2}) [1].

The functional $Z[J]$ is invariant under the field translations
(\ref{label1.6}) if the external source $J(x)$ obeys the constraint
[2]
\begin{eqnarray}\label{label1.8}
\int d^2x\,J(x) = 0.
\end{eqnarray}
In momentum representation
\begin{eqnarray}\label{label1.9}
J(x) = \int \frac{d^2k}{(2\pi)^2}\,\tilde{J}(k)\,e^{\textstyle
-ik\cdot x}
\end{eqnarray}
the constraint (\ref{label1.8}) corresponds to the vanishing of the
Fourier transform of the external source at $k = 0$,
i.e. $\tilde{J}(0) = 0$. This means that (i) the external source
belongs to the class functions obeying the constraint $\tilde{J}(0) =
0$[11,12]\,\footnote{It does not mean that this is Schwartz's class of
test functions ${\cal S}_0(\mathbb{R}^{\,2}) = \{J(x) \in {\cal
S}(\mathbb{R}^{\,2}); \tilde{J}(0) = 0\}$.} and (ii) an external
perturbation does not excite a zero--mode response of the quanta of
the $\vartheta$--field [2].

In Ref.[2] we have shown that the zero--mode of the free massless
(pseudo)scalar field $\vartheta(x)$ describes the motion of the
``center of mass'' of the $\vartheta$--field. This can be easily
comprehended by using a mechanical analog of the free massless
(pseudo)scalar field $\vartheta(x)$ in terms of a chain of $N$ linear
self--coupled oscillators [2,13]. In Ref.[3] we have quantized these
oscillators in terms of the operators of creation and annihilation of
the quanta of the $\vartheta$--field.

Integrating over the $\vartheta$--field in (\ref{label1.4}) we get [2]
\begin{eqnarray}\label{label1.10}
Z[J] = \lim_{\textstyle \mu \to 0}\exp\,\Big\{i\,\frac{1}{2}\int
d^2x\,d^2y\,J(x)\,\Delta(x-y;\mu)\,J(y)\Big\},
\end{eqnarray}
where $\Delta(x-y;\mu)$, the causal two--point Green function, is
related to the two--point Wightman functions [2]
\begin{eqnarray}\label{label1.11}
\Delta(x-y;\mu) &=& i\,\theta(x^0 - y^0)\,D^{(+)}(x - y;\mu) +
i\,\theta(y^0-x^0)\,D^{(-)}(x - y; \mu) =\nonumber\\ &=& -
\frac{i}{4\pi}\,{\ell n}[-\mu^2(x - y)^2 + i\,0].
\end{eqnarray}
Due to the constraint (\ref{label1.8}) the infrared scale $\mu$ can be
replaced by the finite scale $M$ [2]. This yields 
\begin{eqnarray}\label{label1.12}
Z[J] = \exp\,\Big\{i\,\frac{1}{2}\int
d^2x\,d^2y\,J(x)\,\Delta(x-y;M)\,J(y)\Big\},
\end{eqnarray}
where $\Delta(x-y;M)$ is equal to
\begin{eqnarray}\label{label1.13}
\Delta(x-y; M) &=& i\,\theta(x^0 - y^0)\,D^{(+)}(x - y;M) +
i\,\theta(y^0-x^0)\,D^{(-)}(x - y; M) =\nonumber\\ &=& -
\frac{i}{4\pi}\,{\ell n}[- M^2(x - y)^2 + i\,0].
\end{eqnarray}
This gives the Wightman functions $D^{(\pm)}(x; M)$, defined by the
momentum integrals [2]
\begin{eqnarray}\label{label1.14}
\hspace{-0.3in}D^{(\pm)}(x;M)
=\frac{1}{2\pi}\int^{\infty}_{-\infty}\frac{dk^1}{2k^0}
\Big(e^{\textstyle \mp \,i\,k\cdot x} - \cos(k^1\lambda_{\rm
M})\Big),
\end{eqnarray}
where $\lambda_{\rm M} = 1/M$, which are obviously convergent in the
infrared region $k^1 \to 0$ and Lorentz invariant. This solves the
{\it infrared} problem of the free massless (pseudo)scalar field
theory pointed out by Klaiber [10].

Since the free massless (pseudo)scalar field theory free of infrared
divergences describes the bosonized version of the massless Thirring
model in the chirally broken phase, when spontaneous magnetization of
the free massless (pseudo)scalar field defines fully fermion
condensate in the massless Thirring model [1,2], one can expect that
the fermion fields constructed directly from this (pseudo)scalar field
should be equivalent to the Thirring fermion fields quantized in the
chirally broken phase.  

The idea to construct fermion fields in terms of boson fields has been
suggested by Skyrme [14]. The first attempts in this direction have
been undertaken by Lieb and Mattis [15] and Streater and Wilde [16]
within axiomatic quantum field theory. At the standard quantum field
theoretic level the realization of the fermion field operators in
terms of the boson field has been developed by Mandelstam in
connection with the proof of the equivalence between the massive
Thirring model and the sine--Gordon model [17]. Then, it has been
reconsidered and applied to the solution of the massless Thirring
model by different authors [18--20].

The main problem of the construction of fermion fields from the free
massless boson ones [19,20] is in the infrared divergences. The
canonical dimension of a free fermion field is $\sqrt{\mu}$, where
$\mu$ is a typical energy scale.  In the free massless (pseudo)scalar
field theory, suffering with infrared divergences, a typical energy
scale is the infrared cut--off $\mu$, which should be taken in the
limit $\mu \to 0$ in the final expressions. In the boson field
approach to the solution of the massless Thirring model [19,20] the
fermion fields, either implicitly or explicitly, are proportional to
the infrared cut--off $\mu$, $\psi(x) \propto \sqrt{\mu}$, and,
therefore, should vanish in the limit $\mu \to 0$. At an intermediate
step of calculations, when the infrared scale $\mu$ is kept finite but
infinitesimally small, the proportionality of the fermion fields to
the infrared cut--off $\mu$ leads to the violation of the constant of
motion [1]. It is obvious, since the constant does not depend on the
infrared cut--off.

The main aim of this paper is to show that the removal of the infrared
divergences from the quantum field theory of a free massless
(pseudo)scalar field $\vartheta(x)$ described by the Lagrangian
(\ref{label1.5}) [2,3] allows to construct fermion field operators
in terms of the boson fields possessing the properties of the Thirring
fermion fields quantized in the chirally broken phase [1].

The paper is organized as follows. In Section 2 we give a cursory
outline of the massless Thirring model. We write down the Lagrangian
of the self--coupled fermion fields, the equations of motion, the
constant of motion and canonical anti--commutation relations. In
Section 3 we discuss the properties of the free massless
(pseudo)scalar fields defined in the quantum field theory free of
infrared divergences formulated in [2,3]. In Section 4, following
mainly the results obtained by Morchio, Pierotti and Strocchi [20], we
discuss the properties of the free massless fermion fields in the
boson field representation. We show that the fermion field operators
obey all standard canonical anti--commutation relations required for
the free fermion quantum field theory. The self--consistency of all
results is controlled by the normalization scale of the two--point
Wightman functions [2,3]. In Section 5 we discuss the properties of
the massless Thirring fermion fields in the boson field representation
constructed form the boson fields according to the recipe suggested by
Morchio, Pierotti and Strocchi [20]. We show that the fermion field
operators invented by Morchio, Pierotti and Strocchi [20] and
regularized by the finite normalization scale $M$ of the two--point
Wightman functions [2,3] obey all anti--commutation relations and
non--perturbative properties of the Thirring fermion fields [1--4].
In Section 6 we construct the operator of the scalar fermion density
$\bar{\psi}(x)\psi(x)$ in the boson field representation of the
massless Thirring fermion fields discussed in Section 5. We show that
the vacuum expectation value of this operator, i.e. the fermion
condensate, does not vanish. This means that the fermion fields,
constructed from the boson fields, possess the properties of the
massless Thirring fermion fields quantized in the chirally broken
phase. This confirms the results obtained in [1,4] pointing out the
existence of the non--perturbative chirally broken phase in the
massless Thirring model. In Section 7 we show that the Thirring
fermion field operators in the boson field representation obey the
constant of motion revealed in [1]. In the Conclusion we discuss the
obtained results. In the Appendix we formulate the procedure for the
normal ordering of the exponential operators depending on the free
massless boson fields defined by the quantum field theory without
infrared divergences.

\section{Massless Thirring model. Cursory outline}
\setcounter{equation}{0}

\hspace{0.2in} The massless Thirring model [5] is a theory of a
self--coupled Dirac field $\psi(x)$
\begin{eqnarray}\label{label2.1}
{\cal L}_{\rm Th}(x) = \bar{\psi}(x)i\gamma^{\mu}\partial_{\mu}\psi(x) -
\frac{1}{2}\,g\,\bar{\psi}(x)\gamma^{\mu}\psi(x)\bar{\psi}(x)
\gamma_{\mu}\psi(x),
\end{eqnarray}
where $g$ is a dimensionless coupling constant that can be either
positive or negative. Since for $g > 0$ the fermion system described
by the Lagrangian (\ref{label2.1}) is unstable under chiral symmetry
breaking [1--4], we will consider only $g > 0$. 

The field $\psi(x)$ is a spinor field with two components $\psi_1(x)$
and $\psi_2(x)$. The $\gamma$--matrices are defined in terms of the
well--known $2\times 2$ Pauli matrices $\sigma_1$, $\sigma_2$ and
$\sigma_3$
\begin{eqnarray}\label{label2.2}
\gamma^0 = \sigma_1 = \left(\begin{array}{cc}
0 & 1\\
1 & 0 
\end{array}\right)\,,\,\gamma^1 = - i\sigma_2 = \left(\begin{array}{cc}
0 & - 1\\
1 & 0 
\end{array}\right)\,,\,\gamma^5 = \gamma^0\gamma^1 = \sigma_3 = 
\left(\begin{array}{cc} 1 & 0 \\ 0 & -1
\end{array}\right).
\end{eqnarray}
These $\gamma$--matrices obey the standard relations
\begin{eqnarray}\label{label2.3}
\gamma^{\mu}\gamma^{\nu}&+& \gamma^{\nu}\gamma^{\mu} = 2
g^{\mu\nu},\nonumber\\ 
\gamma^{\mu}\gamma^5&+& \gamma^5\gamma^{\mu} = 0.
\end{eqnarray}
We use the metric tensor $g^{\mu\nu}$ defined by $g^{00} = - g^{11} =
1$ and $g^{01} = g^{10} = 0$. The axial--vector product
$\gamma^{\mu}\gamma^5$ can be expressed in terms of $\gamma^{\nu}$
\begin{eqnarray}\label{label2.4}
\gamma^{\mu}\gamma^5 &=& - \epsilon^{\mu\nu}\gamma_{\nu},\
\end{eqnarray}
where $\epsilon^{\mu\nu}$ is the anti--symmetric tensor defined by
$\epsilon^{01} = - \epsilon^{10} = 1$. Further, we also use the
relation $\gamma^{\mu}\gamma^{\nu} = g^{\mu\nu} +
\epsilon^{\mu\nu}\gamma^5$.

The Lagrangian (\ref{label2.1}) is obviously invariant under the
chiral group $U_{\rm V}(1)\times U_{\rm A}(1)$
\begin{eqnarray}\label{label2.5}
\psi(x) \stackrel{\rm V}{\longrightarrow} \psi^{\prime}(x) &=&
e^{\textstyle i\alpha_{\rm V}}\psi(x) ,\nonumber\\ \psi(x)
\stackrel{\rm A}{\longrightarrow} \psi^{\prime}(x) &=& e^{\textstyle
i\alpha_{\rm A}\gamma^5}\psi(x),
\end{eqnarray}
where $\alpha_{\rm V}$ and $\alpha_{\rm A}$ are real parameters
defining global rotations. 

Due to invariance under chiral group $U_{\rm V}(1)\times U_{\rm A}(1)$
the vector and axial--vector current $j^{\mu}(x)$ and $j^{\mu}_5(x)$,
induced by vector (V) and axial--vector (A) rotations and defined by
\begin{eqnarray}\label{label2.6}
j^{\mu}(x) &=&
\bar{\psi}(x)\gamma^{\mu}\psi(x),\nonumber\\
j^{\mu}_5(x) &=&
\bar{\psi}(x)\gamma^{\mu}\gamma^5\psi(x),
\end{eqnarray}
are conserved
\begin{eqnarray}\label{label2.7}
\partial_{\mu}j^{\mu}(x) = \partial_{\mu}j^{\mu}_5(x) = 0.
\end{eqnarray}
Recall, that in 1+1--dimensional field theories the vector and
axial--vector currents are related by $j^{\mu}_5(x) = -
\varepsilon^{\mu\nu}j_{\nu}(x)$ due to the properties of Dirac
matrices (\ref{label2.4}).

Using the Lagrangian (\ref{label2.1}) we derive the equations
of motion
\begin{eqnarray}\label{label2.8}
i\gamma^{\mu}\partial_{\mu}\psi(x) &=&
g\,j^{\mu}(x)\gamma_{\mu}\psi(x) ,\nonumber\\
-i\partial_{\mu}\bar{\psi}(x)\gamma^{\mu} &=&
g\,\bar{\psi}(x,t)\gamma_{\mu}j^{\mu}(x).
\end{eqnarray}
Due to the peculiarity of 1+1--dimensional quantum field theory of the
Thirring fermion fields [1] these equations are equivalent to
\begin{eqnarray}\label{label2.9}
i\partial_{\mu}\psi(x) &=& a\,j_{\mu}(x)\psi(x) +
b\,\varepsilon_{\mu\nu}j^{\nu}(x)\gamma^5\psi(x),\nonumber\\
-i\partial_{\mu}\bar{\psi}(x) &=&a\,\bar{\psi}(x)j_{\mu}(x) +
b\,\bar{\psi}(x)\gamma^5j^{\nu}(x)\varepsilon_{\nu\mu}.
\end{eqnarray}
The parameters $a$ and $b$ are constrained by the condition $a + b =
g$ and $a-b = 1/c$ where $c$ is the Schwinger term [1]. Multiplying
equations (\ref{label2.9}) by $\gamma^{\mu}$ and summing over $\mu =
0,1$ we end up with the equations of motion (\ref{label2.8}). The
Thirring fermion fields evolving according to the equations of motion
(\ref{label2.8}) and (\ref{label2.9}) obey the constant of motion [1]
\begin{eqnarray}\label{label2.10}
[\bar{\psi}(x)\psi(x)]^2 + [\bar{\psi}(x)i\gamma^5\psi(x)]^2 = C,
\end{eqnarray}
where $C$ is a non--vanishing constant. As has been shown in Ref.[1]
$C = M^2_f/g^2$, where $M_f$ is a dynamical mass of the Thirring fermion
fields quantized in the chirally broken phase and defining the fermion
condensate $\langle 0|\bar{\psi}(x)\psi(x)|0\rangle = - M_f/g$
[1,4]. In terms of the components of the fermion field the constant of
motion (\ref{label2.10}) reads
\begin{eqnarray}\label{label2.11}
\psi^{\dagger}_1(x)\psi_2(x)\psi^{\dagger}_2(x)\psi_1(x) +
\psi^{\dagger}_2(x)\psi_1(x)\psi^{\dagger}_1(x)\psi_2(x) =
\frac{1}{2}\,C.
\end{eqnarray}
The conjugate momenta of the fields $\psi_1(x)$ and $\psi_2(x)$ are
$i\psi^{\dagger}_1(x)$ and $i\psi^{\dagger}_2(x)$,
respectively. Therefore, they obey the canonical anti--commutation
relations
\begin{eqnarray}\label{label2.12}
\psi_1(x^0,x^1)\psi^{\dagger}_1(x^0,y^1) +
\psi^{\dagger}_1(x^0,y^1)\psi_1(x^0,x^1) &=& \delta(x^1 -
y^1),\nonumber\\ \psi_2(x^0,x^1)\psi^{\dagger}_2(x^0,y^1) +
\psi^{\dagger}_2(x^0,y^1)\psi_2(x^0,x^1) &=& \delta(x^1 - y^1).
\end{eqnarray}
We would like to emphasize that such canonical anti--commutation
relations are valid for fermion fields with a canonical dimension
$D_{\psi} = 1/2$\,\footnote{The total dimension of the fermion field
$D_{\psi}$ is equal to $D_{\psi} = d_{\psi} + 1/2$, where $d_{\psi}$
is a dynamical dimension [4].}. In 1+1--dimensional space time these
are only free fermion fields. In the case of the Thirring model with
self--coupled fermion fields the total dimension of the fermion fields
differs from the canonical one $D_{\psi} \neq 1/2$ [4]. Therefore, for
the Thirring fermion fields the canonical anti--commutation relation
(\ref{label2.12}) is valid only for {\it bare} fermion fields having
the dimension $D_{\psi} = 1/2$. For the derivation of the canonical
anti--commutation relations for the Thirring fermion fields in the
boson field representation one has to construct the products of the
fermion field operators $\psi_i(x^0,x^1)\psi^{\dagger}_i(x^0,y^1)$ for
$(i=1,2)$ and to subtract the dynamical dimensions by multiplying the
products by the factors $(x^1 - y^1 - i0)^{\textstyle 2D_{\psi}
-1}$. This yields
\begin{eqnarray}\label{label2.13}
\hspace{-0.3in}&&\psi_i(x^0,x^1)\psi^{\dagger}_i(x^0,y^1) \to
\Psi_i(x^0,x^1)\Psi^{\dagger}_i(x^0,y^1) \propto (x^1 - y^1 -
i0)^{\textstyle 2D_{\psi}
-1}\,\psi_i(x^0,x^1)\psi^{\dagger}_i(x^0,y^1).\nonumber\\
\hspace{-0.3in}&&
\end{eqnarray}
For the $\Psi$--fields one would get the canonical equal--time
anti--commutation relations
\begin{eqnarray}\label{label2.14}
\Psi_i(x^0,x^1)\Psi^{\dagger}_i(x^0,y^1) +
\Psi^{\dagger}_i(x^0,y^1)\Psi_i(x^0,x^1) = \delta(x^1 - y^1),\quad
(i=1,2).
\end{eqnarray}
However, considering the spatial coordinates $x^1$ and $y^1$ separated
by an infinitesimal distance, $x^1 \sim y^1$, one can show that the
canonical anti--commutation relations for the massless Thirring
fermions fields $\psi_i(x^0,x^1)$ and $\psi^{\dagger}_i(x^0,y^1)$ can
be written in the form
\begin{eqnarray}\label{label2.15}
\psi_i(x^0,x^1)\psi^{\dagger}_i(x^0,y^1) +
\psi^{\dagger}_i(x^0,y^1)\psi_i(x^0,x^1) =Z_2\, \delta(x^1 -
y^1),\quad (i=1,2).
\end{eqnarray}
where $Z_2$ is the renormalization constant of the wave function of
the massless Thirring fermion fields [4]. The expression
(\ref{label2.15}) agrees with Mandelstam's relation (see Eq.(3.1) of
Ref.[17]).

\section{Free massless boson fields as a basis for the  representation 
of massless fermion fields} 
\setcounter{equation}{0}

\hspace{0.2in} For the construction of fermion fields from the boson
fields we use two free massless local fields $\varphi(x)$ and
$\tilde{\varphi}(x)$ related by\,\footnote{In this paper we will use
as close as it is possible the notations accepted in the paper by
Morchio, Pierotti, and Strocchi [20]. Therefore, the $\varphi$--field
should be an analogy of the $\vartheta$--field.}
\begin{eqnarray}\label{label3.1}
\frac{\partial \varphi(x)}{\partial x^0} = - \frac{\partial
\tilde{\varphi}(x)}{\partial x^1}\quad,\quad \frac{\partial \varphi(x)}{\partial x^1} = - \frac{\partial
\tilde{\varphi}(x)}{\partial x^0}.
\end{eqnarray}
Since the fields $\varphi(x)$ and $\tilde{\varphi}(x)$ describe free
massless (pseudo)scalar fields and their conjugate momenta $\Pi(x) =
\dot{\varphi}(x)$ and $\tilde{\Pi}(x) = \dot{\tilde{\varphi}}(x)$,
they can be represented in the form of expansions into plane waves
[1--3]
\begin{eqnarray}\label{label3.2}
\varphi(x) &=&
\frac{1}{2\pi}\int\limits^{\infty}_{-\infty}\frac{dk^1}{2k^0}\,
\Big[c(k^1) \,e^{\textstyle -ik^0x^0 + ik^1x^1} + c^{\dagger}(k^1)
\,e^{\textstyle +ik^0x^0 - i k^1x^1}\,\Big],\nonumber\\ \Pi(x) &=&
\frac{1}{2\pi}\int\limits^{\infty}_{-\infty}\frac{dk^1}{2i}\,
\Big[c(k^1) \,e^{\textstyle -ik^0x^0 + ik^1x^1} - c^{\dagger}(k^1)
\,e^{\textstyle +ik^0x^0 - i k^1x^1}\,\Big],\nonumber\\
\tilde{\varphi}(x) &=&
\frac{1}{2\pi}\int\limits^{\infty}_{-\infty}\frac{dk^1}{2k^0}\,
\Big[\tilde{c}(k^1) \,e^{\textstyle -ik^0x^0 + ik^1x^1} +
\tilde{c}^{\dagger}(k^1) \,e^{\textstyle +ik^0x^0 - i
k^1x^1}\,\Big],\nonumber\\ \tilde{\Pi}(x) &=&
\frac{1}{2\pi}\int\limits^{\infty}_{-\infty}\frac{dk^1}{2i}\,
\Big[\tilde{c}(k^1) \,e^{\textstyle -ik^0x^0 + ik^1x^1} -
\tilde{c}^{\dagger}(k^1) \,e^{\textstyle +ik^0x^0 - i k^1x^1}\,\Big].
\end{eqnarray}
The operators of annihilation and creation of quanta of the fields
$\varphi(x)$ and $\tilde{\varphi}(x)$ obey standard commutation
relations
\begin{eqnarray}\label{label3.3}
&&[c(k^1), c^{\dagger}(q^1)] = (2\pi)\,2k^0\,\delta(k^1 -
q^1),\nonumber\\ &&[\tilde{c}(k^1), \tilde{c}^{\dagger}(q^1)] =
(2\pi)\,2k^0\,\delta(k^1 - q^1),
\end{eqnarray}
which give the equal--time commutation relations
\begin{eqnarray}\label{label3.4}
&&[\varphi(x^0,x^1),\Pi(x^0,y^1)] = i\,\delta(x^1 - y^1),\nonumber\\
&&[\tilde{\varphi}(x^0,x^1),\tilde{\Pi}(x^0,y^1)] = i\,\delta(x^1 -
y^1).
\end{eqnarray}
According to the definition (\ref{label3.1}) the operators $c(k^1)$
and $\tilde{c}(k^1)$ are related by
\begin{eqnarray}\label{label3.5}
k^1\,c(k^1) = k^0\,\tilde{c}(k^1)\quad,\quad k^1\,c^{\dagger}(k^1) =
k^0\,\tilde{c}^{\dagger}(k^1).
\end{eqnarray}
These relations can be rewritten in terms of the sign function
$\varepsilon(k^1)$
\begin{eqnarray}\label{label3.6}
\tilde{c}(k^1) = \varepsilon(k^1)\,c(k^1)\quad,\quad
\tilde{c}^{\dagger}(k^1) = \varepsilon(k^1)\,c^{\dagger}(k^1).
\end{eqnarray}
The fields $\varphi(x)$ and $\tilde{\varphi}(x)$ are supplemented by
the charges $Q(x^0)$ and $\tilde{Q}(x^0)$
\begin{eqnarray}\label{label3.7}
Q(x^0) &=& \int^{\infty}_{-\infty}dx^1\,\Pi(x^0,x^1),\nonumber\\
\tilde{Q}(x^0) &=& \int^{\infty}_{-\infty}dx^1\,\tilde{\Pi}(x^0,x^1).
\end{eqnarray}
We suggest to regularize these charges according to the procedure
developed in Refs.[1--3] and get
\begin{eqnarray}\label{label3.8}
\hspace{-0.3in}&&Q(x^0) = \lim_{V \to \infty}Q(x^0;V)=\nonumber\\ 
\hspace{-0.3in}&&=\lim_{V\to
\infty}\frac{-iV}{4\sqrt{\pi}}\int^{\infty}_{-\infty}dk^1\,
[c(k^1)\,e^{\textstyle -i|k^1|x^0} - c^{\dagger}(k^1)\,e^{\textstyle
+i|k^1|x^0}\,]\,e^{\textstyle -V^2(k^1)^2/4}\,,\nonumber\\
\hspace{-0.3in}&&\tilde{Q}(x^0) =\lim_{V \to
\infty}\tilde{Q}(x^0;V)=\nonumber\\
\hspace{-0.3in}&&= \lim_{V\to
\infty}\frac{-iV}{4\sqrt{\pi}}\int^{\infty}_{-\infty}dk^1\,
[\tilde{c}(k^1)\,e^{\textstyle -i|k^1|x^0} -
\tilde{c}^{\dagger}(k^1)\,e^{\textstyle +i|k^1|x^0}\,]\,e^{\textstyle
-V^2(k^1)^2/4},
\end{eqnarray}
where $V$ is the spatial extent of the system. The charge operators
$Q(x^0)$ and $\tilde{Q}(x^0)$ are the generators of a $U(1)\times
U(1)$ symmetry related to the shifts of the fields $\varphi(x)$ and
$\tilde{\varphi}(x)$
\begin{eqnarray}\label{label3.9}
\varphi(x) &\to& \varphi\,'(x)\stackrel{Q}{=} \varphi(x) +
\alpha,\nonumber\\ \tilde{\varphi}(x) &\to&
\tilde{\varphi}\,'(x)\stackrel{Q}{=} \tilde{\varphi}(x),\nonumber\\
\varphi(x) &\to& \varphi\,'(x)\stackrel{\tilde{Q}}{=} \varphi(x)
,\nonumber\\ \tilde{\varphi}(x) &\to&
\tilde{\varphi}\,'(x)\stackrel{\tilde{Q}}{=} \tilde{\varphi}(x) +\beta,
\end{eqnarray}
where $\alpha$ and $\beta$ are the group parameters $\alpha, \beta \in
\mathbb{R}^1$. As we have shown in Refs.[1--3] these symmetries are
spontaneously broken and the Goldstone bosons are the quanta of the
fields $\varphi(x)$ and $\tilde{\varphi}(x)$.

For the construction of fermion field operators from the boson
operators it is convenient to introduce {\it left} and {\it right}
fields $\varphi_L(x)$ and $\varphi_R(x)$ defined by [20]
\begin{eqnarray}\label{label3.10}
\varphi_L(x) = \frac{1}{2}\,[\varphi(x) -
\tilde{\varphi}(x)]\quad,\quad\varphi_R(x) =
\frac{1}{2}\,[\varphi(x) + \tilde{\varphi}(x)],
\end{eqnarray}
which commute $[\varphi_L(x), \varphi_R(y)] = 0$ for arbitrary
space--time points $x$ and $y$.

The expansions of these  fields and their conjugate momenta $\Pi_L(x)$
and $\Pi_R(x)$ into plane waves read
\begin{eqnarray}\label{label3.11}
\hspace{-0.3in}\varphi_L(x) &=&
\frac{1}{2\pi}\int\limits^{\infty}_{-\infty}\frac{dk^1}{2k^0}\,
\theta(-k^1)\, \Big[c(k^1) \,e^{\textstyle -ik^0x^0 + ik^1x^1} +
c^{\dagger}(k^1) \,e^{\textstyle +ik^0x^0 - i
k^1x^1}\,\Big],\nonumber\\ \hspace{-0.3in}\Pi_L(x) &=&
\frac{1}{2\pi}\int\limits^{\infty}_{-\infty}\frac{dk^1}{2i}\,
\theta(-k^1)\, \Big[c(k^1) \,e^{\textstyle -ik^0x^0 + ik^1x^1} -
c^{\dagger}(k^1) \,e^{\textstyle +ik^0x^0 - i
k^1x^1}\,\Big],\nonumber\\ \hspace{-0.3in}\varphi_R(x) &=&
\frac{1}{2\pi}\int\limits^{\infty}_{-\infty}\frac{dk^1}{2k^0}\,
\theta(+k^1)\, \Big[c(k^1) \,e^{\textstyle -ik^0x^0 + ik^1x^1} +
c^{\dagger}(k^1) \,e^{\textstyle +ik^0x^0 - i
k^1x^1}\,\Big],\nonumber\\ \hspace{-0.3in}\Pi_R(x) &=&
\frac{1}{2\pi}\int\limits^{\infty}_{-\infty}\frac{dk^1}{2i}\,
\theta(+k^1)\, \Big[c(k^1) \,e^{\textstyle -ik^0x^0 + ik^1x^1} -
c^{\dagger}(k^1) \,e^{\textstyle +ik^0x^0 - i k^1x^1}\,\Big],
\end{eqnarray}
where $\theta(\pm k^1)$ are the Heaviside functions. The charge
operators $Q_L(x^0)$ and $Q_R(x^0)$ are defined by
\begin{eqnarray}\label{label3.12}
\hspace{-0.5in}&&Q_L(x^0) = \lim_{V \to \infty}Q_L(x^0; V)
=\nonumber\\
\hspace{-0.5in}&&= \lim_{V\to
\infty}\frac{-iV}{\sqrt{\pi}}\int^{\infty}_{-\infty}dk^1\,
\theta(-k^1)\,
[c(k^1)\,e^{\textstyle -i|k^1|x^0} - c^{\dagger}(k^1)\,e^{\textstyle
+i|k^1|x^0}\,]\,e^{\textstyle -V^2(k^1)^2/4}\,,\nonumber\\
\hspace{-0.5in}&&Q_R(x^0) = \lim_{V \to \infty}Q_R(x^0; V)
=\nonumber\\
\hspace{-0.5in}&&= \lim_{V\to
\infty}\frac{-iV}{2\sqrt{\pi}}\int^{\infty}_{-\infty}dk^1\,
\theta(+k^1)\, [c(k^1)\,e^{\textstyle -i|k^1|x^0} -
c^{\dagger}(k^1)\,e^{\textstyle +i|k^1|x^0}\,]\,e^{\textstyle
-V^2(k^1)^2/4}.
\end{eqnarray}
The fields $\varphi_L(x)$ and $\varphi_R(x)$ and their charge
operators have the following commutation relations [20]
\begin{eqnarray}\label{label3.13}
[\varphi_L(x),Q_L(x^0)] = [\varphi_R(x),Q_R(x^0)]
=[\varphi_L(x),Q_L(0)] = [\varphi_R(x),Q_R(0)] = -i.
\end{eqnarray}
The two--point Wightman functions of the fields $\varphi_L(x)$ and
$\varphi_R(x)$ are equal to
\begin{eqnarray}\label{label3.14}
D^{(+)}_L(x-y; M) = \langle 0|\varphi_L(x)\varphi_L(y)|0\rangle = -
\frac{1}{4\pi}\,{\ell n}[iM(x_+ - y_+ - i0)],\nonumber\\
D^{(+)}_R(x-y; M) = \langle 0|\varphi_R(x)\varphi_R(y)|0\rangle = -
\frac{1}{4\pi}\,{\ell n}[iM(x_- - y_- - i0)],
\end{eqnarray}
where we have denoted $x_{\pm} = x^0 \pm x^1$ and $y_{\pm} = y^0 \pm
y^1$, and $M$ is a finite normalization scale [2,3].  The sum of the
Wightman functions (\ref{label3.14}) is equal to $D^{(+)}(x-y; M)$
\begin{eqnarray}\label{label3.15}
&&D^{(+)}(x-y; M) = D^{(+)}_L(x-y; M) + D^{(+)}_R(x-y; M) = -
\frac{1}{4\pi}\,{\ell n}[iM(x_+ - y_+ - i0)]\nonumber\\ && -
\frac{1}{4\pi}\,{\ell n}[iM(x_- - y_- - i0)] = -\frac{1}{4\pi}\,{\ell
n}[- M^2(x-y)^2 +i0\cdot\varepsilon(x^0-y^0)].
\end{eqnarray}
In turn, the difference of the Wightman functions (\ref{label3.14})
does not depend on $M$ and amounts to
\begin{eqnarray}\label{label3.16}
D^{(+)}_L(x-y; M) - D^{(+)}_R(x-y; M) = - \frac{1}{4\pi}\,{\ell
n}\Big(\frac{x_+ - y_+ - i0}{x_- - y_- - i0}\Big).
\end{eqnarray}
The r.h.s. of (\ref{label3.16}) is the well--known Wightman function
$D^{(+)}_5(x-y; M)$ defined by [4]
\begin{eqnarray}\label{label3.17}
D^{(+)}_5(x-y; M) = \langle 0|\varphi(x)\tilde{\varphi}(y)|0\rangle
=\langle 0|\tilde{\varphi}(x)\varphi(y)|0\rangle = -
\frac{1}{4\pi}\,{\ell n}\Big(\frac{x_+ - y_+ - i0}{x_- - y_- -
i0}\Big).
\end{eqnarray}
Therefore, for further analysis we will denote
\begin{eqnarray}\label{label3.18}
D^{(+)}_L(x-y; M) - D^{(+)}_R(x-y; M) = D^{(+)}_5(x-y; M).
\end{eqnarray}
Now we are able to construct the fermion fields from the free massless
boson fields $\varphi_L(x)$, $\varphi_R(x)$ and their charges $Q_L(0)$
and $Q_R(0)$.

\section{Free fermion fields in the boson field representation}
\setcounter{equation}{0}

\hspace{0.2in} Following Morchio, Pierotti and Strocchi [20] we
represent the free massless fermion fields $\Psi_1(x)$ and $\Psi_2(x)$
in terms of the fields $\varphi_L(x)$ and $\varphi_R(x)$ and their
charge operators $Q_L(0)$ and $Q_R(0)$
\begin{eqnarray}\label{label4.1}
\Psi_1(x^0,x^1) &=& + \,i\sqrt{\frac{M}{2\pi}}\,e^{\textstyle -
i(\sqrt{\pi}/4)Q_R(0)}:\!e^{\textstyle
2i\sqrt{\pi}\,\varphi_L(x^0,x^1)}\!:,\nonumber\\ \Psi_2(x^0,x^1) &=&
-\,i\sqrt{\frac{M}{2\pi}}\,e^{\textstyle +
i(\sqrt{\pi}/4)Q_L(0)}:\!e^{\textstyle
2i\sqrt{\pi}\,\varphi_R(x^0,x^1)}\!:,
\end{eqnarray}
where $M$ is the normalization scale in the Wightman functions
(\ref{label3.14}).

The anti--commutativity of the fermion field operators can be verified
by analysing the products $\Psi_i(x)\Psi_j(y)$ and
$\Psi_i(x)\Psi^{\dagger}_j(y)$. For the product
$\Psi_1(x^0,x^1)\Psi_1(y^0,y^1)$ we get\,\footnote{In the Appendix we
formulate the procedure for the normal ordering of the exponential
operators depending on the free massless boson fields defined in the
quantum field theory without infrared divergences.}
\begin{eqnarray}\label{label4.2}
\hspace{-0.5in}&&\Psi_1(x^0,x^1)\Psi_1(y^0,y^1) = -\,
\frac{M}{2\pi}\,e^{\textstyle - i(\sqrt{\pi}/4)Q_R(0)}:\!e^{\textstyle
2i\sqrt{\pi}\,\varphi_L(x^0,x^1)}\!:e^{\textstyle -
i(\sqrt{\pi}/4)Q_R(0)}\nonumber\\
\hspace{-0.5in}&&\times\,:\!e^{\textstyle
2i\sqrt{\pi}\,\varphi_L(y^0,y^1)}\!: = -
\,\frac{M}{2\pi}\,e^{\textstyle -
i(\sqrt{\pi}/2)Q_R(0)}:\!e^{\textstyle
2i\sqrt{\pi}\,\varphi_L(x^0,x^1)}\!::\!e^{\textstyle
2i\sqrt{\pi}\,\varphi_L(y^0,y^1)}\!:=\nonumber\\\hspace{-0.5in} &&=
e^{\textstyle
-4\pi\,[\varphi^{(+)}_L(x^0,x^1),\varphi^{(-)}_L(y^0,y^1)]}\nonumber\\
\hspace{-0.5in}&&\times\,(-1)\,\frac{M}{2\pi}\,e^{\textstyle -
i(\sqrt{\pi}/2)Q_R(0)} :\!e^{\textstyle
2i\sqrt{\pi}\,(\varphi_L(x^0,x^1) + \varphi_L(y^0,y^1))}\!:,
\end{eqnarray}
where we have used the relation [17]
\begin{eqnarray}\label{label4.3}
&&:\!e^{\textstyle
2i\sqrt{\pi}\,\varphi_L(x^0,x^1)}\!::\!e^{\textstyle 2i\sqrt{\pi}\,
\varphi_L(y^0,y^1)}\!: = e^{\textstyle -4\pi\,
[\varphi^{(+)}_L(x^0,x^1),\varphi^{(-)}_L(y^0,y^1)]}\nonumber\\
&&\times:\!e^{\textstyle 2i\sqrt{\pi}\,(\varphi_L(x^0,x^1) +
\varphi_L(y^0,y^1))}\!:.
\end{eqnarray}
The commutator $[\varphi^{(+)}_L(x^0,x^1),\varphi^{(-)}_L(y^0,y^1)]$
is the Wightman function $D^{(+)}_L(x-y; M)$
\begin{eqnarray}\label{label4.4}
 D^{(+)}_L(x-y; M)=
[\varphi^{(+)}_L(x^0,x^1),\varphi^{(-)}_L(y^0,y^1)] = \langle
0|\varphi_L(x^0,x^1)\varphi_L(y^0,y^1)|0\rangle.
\end{eqnarray}
Thus, we get (see the Appendix)
\begin{eqnarray}\label{label4.5}
-4\pi\,[\varphi^{(+)}_L(x^0,x^1),\varphi^{(-)}_L(y^0,y^1)] = {\ell
 n}[iM\,(x_+ - y_+ - i0)].
\end{eqnarray}
Substituting (\ref{label4.5}) in (\ref{label4.2}) we obtain
\begin{eqnarray}\label{label4.6}
\hspace{-0.3in}&&\Psi_1(x^0,x^1)\Psi_1(y^0,y^1) = -
\Psi_1(y^0,y^1)\Psi_1(x^0,x^1)=\nonumber\\
\hspace{-0.3in}&&=\frac{M^2}{2\pi i}\,(x_+ - y_+ - i0)\,e^{\textstyle
- i(\sqrt{\pi}/2)Q_R(0)} :\!e^{\textstyle
2i\sqrt{\pi}\,(\varphi_L(x^0,x^1) + \varphi_L(y^0,y^1))}\!:.
\end{eqnarray}
Thus, the operators $\Psi_1(x^0,x^1)$ and $\Psi_1(y^0,y^1)$ obey the
anti--commutation relation. For $x = y$ the product of the fermion
fields $\Psi^2_1(x^0,x^1) = 0$ vanishes as required for fermion fields
[5].

Now let us show that the fields $\Psi_1(x^0,x^1)$ and
$\Psi^{\dagger}_1(x^0,y^1)$ satisfy the canonical anti--commutation
relations (\ref{label2.12}). We get
\begin{eqnarray}\label{label4.7}
\hspace{-0.3in}&&\Psi_1(x^0,x^1)\Psi^{\dagger}_1(x^0,y^1) =
\frac{M}{2\pi}\,e^{\textstyle - i(\sqrt{\pi}/4)Q_R(0)}:\!e^{\textstyle
+ 2i\sqrt{\pi}\,\varphi_L(x^0,x^1)}\!:e^{\textstyle +
i(\sqrt{\pi}/4)Q_R(0)}\nonumber\\
\hspace{-0.3in}&&\times\,:\!e^{\textstyle -
2i\sqrt{\pi}\,\varphi_L(x^0,y^1)}\!: = \nonumber\\
\hspace{-0.3in}&&=\frac{M}{2\pi}\,e^{\textstyle +
4\pi\,[\varphi^{(+)}_L(x^0,x^1),\varphi^{(-)}_L(x^0,y^1)]}
:\!e^{\textstyle + 2i\sqrt{\pi}\,[\varphi_L(x^0,x^1) -
\varphi_L(x^0,y^1)]}\!: =\nonumber\\ \hspace{-0.3in}&&=
\frac{M}{2\pi}\,e^{\textstyle - {\ell n}[- i\,M\,(x^1 - y^1 +
i0)]}:\!e^{\textstyle + 2i\sqrt{\pi}\,[\varphi_L(x^0,x^1) -
\varphi_L(x^0,y^1)]}\!: =\nonumber\\
\hspace{-0.3in}&&= \frac{i}{2\pi}\,\frac{1}{x^1 - y^1 +
i0}\,:\!e^{\textstyle + 2i\sqrt{\pi}\,[\varphi_L(x^0,x^1) -
\varphi_L(x^0,y^1)]}\!:.
\end{eqnarray}
For the product $\Psi^{\dagger}_1(x^0,y^1)\Psi_1(x^0,x^1)$ we get
\begin{eqnarray}\label{label4.8}
\Psi^{\dagger}_1(x^0,y^1)\Psi_1(x^0,x^1) =
\frac{i}{2\pi}\,\frac{1}{y^1 - x^1 + i0}:\!e^{\textstyle +
2i\sqrt{\pi}\,[\varphi_L(x^0,x^1) - \varphi_L(x^0,y^1)]}\!:.
\end{eqnarray}
Using relations (\ref{label4.7}) and (\ref{label4.8}) we derive the
canonical anti--commutation relation
\begin{eqnarray}\label{label4.9}
&&\{\Psi_1(x^0,x^1), \Psi^{\dagger}_1(x^0,y^1)\} =
\frac{i}{2\pi}\Big[\frac{1}{x^1 - y^1 + i0} - \frac{1}{x^1 - y^1 -
i0}\Big]\nonumber\\ &&\times\,:\!e^{\textstyle +
2i\sqrt{\pi}\,[\varphi_L(x^0,x^1) -
\varphi_L(x^0,y^1)]}\!: = \delta(x^1 - y^1).
\end{eqnarray}
The other standard anti--commutation relations
\begin{eqnarray}\label{label4.10}
&&\{\Psi_1(x^0,x^1),\Psi_2(y^0,y^1)\}= 0,\nonumber\\
&&\{\Psi_2(x^0,x^1),\Psi_2(y^0,y^1)\}= 0,\nonumber\\
&&\{\Psi_2(x^0,x^1), \Psi^{\dagger}_2(x^0,y^1)\} = \delta(x^1 - y^1)
\end{eqnarray}
can be derived in a similar way. 

We would like to emphasize that the canonical anti--commutation
relations (\ref{label4.9}) and (\ref{label4.10}), proportional to the
$\delta$--function $\delta(x^1-y^1)$, can be derived within the boson
field representation only for a free fermion field with a canonical
dimension $D_{\psi} = 1/2$. In the case of the massless Thirring
model, where the fermion fields have a total dimension $D_{\psi} \neq
1/2$, the derivation of the canonical anti--commutation relations like
(\ref{label4.9}) can be obtained in the boson representation of the
massless Thirring fermion fields only after the subtracting the
dynamical dimension [4].

\section{Massless Thirring fermion fields in the boson field
representation}
\setcounter{equation}{0}

\hspace{0.2in} The free massless Thirring fermion fields $\psi_1(x)$
and $\psi_2(x)$ in the boson field representation [20] are defined by
\begin{eqnarray}\label{label5.1}
\hspace{-0.5in}\psi_1(x) &=& + i\sqrt{\frac{M}{2\pi}}\,e^{\textstyle
-i(\pi/4)(b^{-1}Q_L(0) +
a^{-1}Q_R(0))}:\!e^{\textstyle
+i(a\,\varphi_L(x) + b\,\varphi_R(x))}\!:,\nonumber\\
\hspace{-0.5in}\psi_2(x) &=& - i \sqrt{\frac{M}{2\pi}}\,e^{\textstyle
+i (\pi/4) (a^{-1} Q_L(0) + b^{-1} Q_R(0))}:\!e^{\textstyle
+i(b\,\varphi_L(x) + a\,\varphi_R(x))}\!:.
\end{eqnarray}
The parameters $a$ and $b$ will be adjusted to fulfill the constraints
imposed by the anti--commutativity of the Thirring fermion field
operators $\psi_1(x)$ and $\psi_2(x)$ and their conjugate momenta.

For the verification of anti--commutativity of the fermion operators
and the canonical anti--commutation relations (\ref{label2.12}) we
have to consider the products of the fermion field operators
$\psi_i(x)\psi_j(y)$ and $\psi^{\dagger}_i(x)\psi_i(y)$ for $(i, j =
1,2)$.

In terms of the fermion field operators (\ref{label5.1}) the product
$\psi_1(x)\psi_1(y)$ is defined by
\begin{eqnarray}\label{label5.2}
\hspace{-0.5in}&&\psi_1(x)\psi_1(y) =
-\,\frac{M}{2\pi}\,e^{\textstyle
-i(\pi/4)(b^{-1}Q_L(0) +
a^{-1}Q_R(0))}:\!e^{\textstyle
+i(a\,\varphi_L(x) + b\,\varphi_R(x))}\!:\nonumber\\
\hspace{-0.5in}&&\times\,e^{\textstyle -i(\pi/4)(b^{-1}Q_L(0) +
a^{-1}Q_R(0))}:\!e^{\textstyle +i(a\,\varphi_L(y) +
b\,\varphi_R(y))}\!: = \nonumber\\
\hspace{-0.5in}&&= -\,\frac{M}{2\pi}\,\,e^{\textstyle
-i(\pi/2)(b^{-1}Q_L(0) + a^{-1}Q_R(0))}\nonumber\\
\hspace{-0.5in}&&\times\,e^{\textstyle (\pi/4)\,
[a\,\varphi_L(x) + b\,\varphi_R(x), b^{-1}Q_L(0) 
+ a^{-1}Q_R(0)]}\nonumber\\
\hspace{-0.5in}&&\times\,e^{\textstyle -\,[a\,\varphi^{(+)}_L(x) +
b\,\varphi^{(+)}_R(x),a\,\varphi^{(-)}_L(y) 
+ b\,\varphi^{(-)}_R(y)]}\nonumber\\
\hspace{-0.5in}&&\times\,:\!e^{\textstyle +i a\,(\varphi_L(x) +
\varphi_L(y)) + i b\,(\varphi_R(x) + \varphi_R(y))}\!:,
\end{eqnarray}
where we have used the relations [17] 
\begin{eqnarray}\label{label5.3}
e^{\textstyle\,A}\,e^{\textstyle \,B} = e^{\textstyle \,
[A,B]}\,e^{\textstyle
\,B}\,e^{\textstyle\,A}\quad,\quad:\!e^{\textstyle\,A}\!:
:\!e^{\textstyle \,B}\!: = e^{\textstyle \,
[A^{(+)},B^{(-)}]}\,:\!e^{\textstyle \,A + B}\!:.
\end{eqnarray}
The commutators $[\varphi^{(+)}_L(x),\varphi^{(-)}_L(y)]$ and
$[\varphi^{(+)}_R(x),\varphi^{(-)}_R(y)]$ are defined by the Wightman
functions $D^{(+)}_L(x-y; M)$ and $D^{(+)}_R(x-y; M)$ (see the
Appendix)
\begin{eqnarray}\label{label5.4}
\hspace{-0.5in}&& D^{(+)}_L(x-y; M) =
[\varphi^{(+)}_L(x),\varphi^{(-)}_L(y)] = \langle
0|\varphi_L(x)\varphi_L(y)|0\rangle = -\frac{1}{4\pi}{\ell n}[iM\,(x_+
- y_+ - i0)],\nonumber\\\hspace{-0.5in}&& D^{(+)}_R(x-y; M) =
[\varphi^{(+)}_R(x),\varphi^{(-)}_R(y)] = \langle
0|\varphi_R(x)\varphi_R(y)|0\rangle = -\frac{1}{4\pi}{\ell n}[iM\,(x_-
- y_- - i0)]\nonumber\\\hspace{-0.5in}&&
\end{eqnarray}
and 
\begin{eqnarray}\label{label5.5}
[a\,\varphi_L(x) + b\,\varphi_R(x), b^{-1}Q_L(0) 
+ a^{-1}Q_R(0)] = -i\,\Big(\frac{a}{b} + \frac{b}{a}\,\Big).
\end{eqnarray}
Substituting (\ref{label5.4}) and (\ref{label5.5}) in (\ref{label5.2})
we get
\begin{eqnarray}\label{label5.6}
\hspace{-0.5in}&&\psi_1(x)\psi_1(y) = -\,\frac{M}{2\pi}\,e^{\textstyle
-i\pi\,(a^2 + b^2)/4ab}\,e^{\textstyle -i(\pi/2)(b^{-1}Q_L(0) +
a^{-1}Q_R(0))}\nonumber\\
\hspace{-0.5in}&&\times\,[iM\,(x_+ - y_+ - i0)]^{\textstyle
\,a^2/4\pi}[iM\,(x_- - y_- - i0)]^{\textstyle \,b^2/4\pi}\nonumber\\
\hspace{-0.5in}&&\times\,:\!e^{\textstyle +ia\,(\varphi_L(x) +
\varphi_L(y)) + ib\,(\varphi_R(x) + \varphi_R(y))}\!:.
\end{eqnarray}
The important property of this expression is its vanishing for $x =
y$, $\psi^2_1(x) = 0$, as it is required for fermion fields.

Now let us verify the anti--commutativity of the operators $\psi_1(x)$
and $\psi_1(y)$. For this aim we represent the r.h.s. of
(\ref{label5.2}) as follows
\begin{eqnarray}\label{label5.7}
\hspace{-0.5in}&&\psi_1(x)\psi_1(y) = e^{\textstyle (\pi/4)\,
[a\,\varphi_L(x) + b\,\varphi_R(x), b^{-1}Q_L(0) + a^{-1}Q_R(0)]}
\nonumber\\
\hspace{-0.5in}&&\times\,e^{\textstyle -\,[a\,\varphi^{(+)}_L(x) +
b\,\varphi^{(+)}_R(x),a\,\varphi^{(-)}_L(y) +
b\,\varphi^{(-)}_R(y)]}\nonumber\\
\hspace{-0.5in}&&\times\,e^{\textstyle (\pi/4)\,[b^{-1}Q_L(0) +
a^{-1}Q_R(0), a\,\varphi_L(y) + b\,\varphi_R(y)]}\nonumber\\
\hspace{-0.5in}&&\times\,e^{\textstyle -\,[a\,\varphi_L(y) +
b\,\varphi_R(y),a\,\varphi_L(x) +
b\,\varphi_R(x)]}\,\psi_1(y)\psi_1(x) = \nonumber\\
\hspace{-0.5in}&&= e^{\textstyle -\,a^2(D^{(+)}_L(x-y; M) -
D^{(+)}_L(y-x; M)) - \,b^2(D^{(+)}_R(x-y; M) - D^{(+)}_R(y-x;
M))}\nonumber\\
\hspace{-0.5in}&&\times\, \psi_1(y)\psi_1(x).
\end{eqnarray}
Since the differences of the Wightman functions are equal to
\begin{eqnarray}\label{label5.8}
D^{(+)}_L(x-y; M) - D^{(+)}_L(y-x; M) &=& \frac{i}{4}\,\varepsilon(x_+
- y_+),\nonumber\\ D^{(+)}_R(x-y; M) - D^{(+)}_R(y-x; M) &=&
\frac{i}{4}\,\varepsilon(x_- - y_-),
\end{eqnarray}
substituting (\ref{label5.8}) in (\ref{label5.7}) we obtain
\begin{eqnarray}\label{label5.9}
\hspace{-0.5in}&&\psi_1(x)\psi_1(y) = e^{\textstyle
+\,i\,(a^2/4)\,\varepsilon(x_+ - y_+) + i \,(b^2/4)\,\varepsilon(x_- -
y_-)}\,\psi_1(y)\psi_1(x).
\end{eqnarray}
This expression agrees fully with the result obtained by Morchio,
Pierotti and Strocchi [20]. According to Morchio, Pierotti and
Strocchi [20] the exponential in (\ref{label5.9}) can get $(-1)$ for a
certain choice of the parameters $a$ and $b$.

Now let us show that the fields $\psi_1(x^0,x^1)$ and
$\psi^{\dagger}_1(x^0,y^1)$ satisfy the canonical anti--commutation
relations (\ref{label2.12}). The product of the operators
$\psi_1(x^0,x^1)\psi^{\dagger}_1(x^0,y^1)$ is equal to
\begin{eqnarray}\label{label5.10}
\hspace{-0.3in}&&\psi_1(x^0,x^1)\psi^{\dagger}_1(x^0,y^1) =\nonumber\\
\hspace{-0.5in}&&= \frac{M}{2\pi}\,e^{\textstyle
-i(\pi/4)(b^{-1}Q_L(0) + a^{-1}Q_R(0))}:\!e^{\textstyle
+i(a\,\varphi_L(x^0, x^1) + b\,\varphi_R(x^0, x^1))}\!:\nonumber\\
\hspace{-0.5in}&&\times:\!e^{\textstyle - i(a\,\varphi_L(x^0,y^1) +
b\,\varphi_R(x^0,y^1))}\!:e^{\textstyle +i(\pi/4)(b^{-1}Q_L(0) +
a^{-1}Q_R(0))}=\nonumber\\
\hspace{-0.5in}&&=\frac{M}{2\pi}\,e^{\textstyle + a^2\,D^{(+)}_L(x^1 -
y^1; M) + b^2\,D^{(+)}_R(x^1 - y^1; M)}\,e^{\textstyle
-i(\pi/4)(b^{-1}Q_L(0) + a^{-1}Q_R(0))}\nonumber\\
\hspace{-0.5in}&&\times\,:\!e^{\textstyle +i(a\,\varphi_L(x^0, x^1) +
b\,\varphi_R(x^0, x^1)) - i(a\,\varphi_L(x^0,y^1) +
b\,\varphi_R(x^0,y^1))}\!:\nonumber\\
\hspace{-0.5in}&&\times\,e^{\textstyle +i(\pi/4)(b^{-1}Q_L(0) +
a^{-1}Q_R(0))}=\frac{M}{2\pi}\,e^{\textstyle + a^2\,D^{(+)}_L(x^1 -
y^1; M) + b^2\,D^{(+)}_R(x^1 - y^1; M)}\nonumber\\
\hspace{-0.5in}&&\times\,:\!e^{\textstyle +i(a\,\varphi_L(x^0, x^1) +
b\,\varphi_R(x^0, x^1)) - i(a\,\varphi_L(x^0,y^1) +
b\,\varphi_R(x^0,y^1))}\!:.
\end{eqnarray}
For the product $\psi^{\dagger}_1(x^0,y^1)\psi_1(x^0,x^1)$ we get
\begin{eqnarray}\label{label5.11}
\hspace{-0.3in}&&\psi^{\dagger}_1(x^0,y^1)\psi_1(x^0,x^1) =
\frac{M}{2\pi}\,e^{\textstyle + a^2\,D^{(+)}_L(y^1 - x^1; M) +
b^2\,D^{(+)}_R(y^1 - x^1; M)}\nonumber\\
\hspace{-0.5in}&&\times\,:\!e^{\textstyle +i(a\,\varphi_L(x^0, x^1) +
b\,\varphi_R(x^0, x^1)) - i(a\,\varphi_L(x^0,y^1) +
b\,\varphi_R(x^0,y^1))}\!:.
\end{eqnarray}
Hence, the equal--time anti--commutation relation reads
\begin{eqnarray}\label{label5.12}
\hspace{-0.3in}&&\{\psi_1(x^0,x^1),\psi^{\dagger}_1(x^0,y^1)\} =
\psi_1(x^0,x^1)\psi^{\dagger}_1(x^0,y^1) +
\psi^{\dagger}_1(x^0,y^1)\psi_1(x^0,x^1) = \nonumber\\
\hspace{-0.5in}&&= \frac{M}{2\pi}\,\Bigg[e^{\textstyle +
a^2\,D^{(+)}_L(x^1 - y^1; M) + b^2\,D^{(+)}_R(x^1 - y^1; M)}\nonumber\\
\hspace{-0.5in}&& + e^{\textstyle + a^2\,D^{(+)}_L(y^1 - x^1; M) +
b^2\,D^{(+)}_R(y^1 - x^1; M)}\,\Bigg]\nonumber\\
\hspace{-0.5in}&&\times\,:\!e^{\textstyle +i(a\,\varphi_L(x^0, x^1) +
b\,\varphi_R(x^0, x^1)) - i(a\,\varphi_L(x^0,y^1) +
b\,\varphi_R(x^0,y^1))}\!:.
\end{eqnarray}
The Wightman functions give the following contribution
\begin{eqnarray}\label{label5.13}
&& a^2\,D^{(+)}_L(x^1 - y^1; M) + b^2\,D^{(+)}_R(x^1 - y^1; M)
=\nonumber\\ &&= -\frac{a^2}{4\pi}\,{\ell n}[iM(x^1 - y^1 - i0)] -
\frac{b^2}{4\pi}\,{\ell n}[- iM(x^1 - y^1 - i0)] =\nonumber\\ &&= -
i\,\frac{a^2 - b^2}{8} - \frac{a^2 + b^2}{4\pi}\,{\ell n}[M(x^1 - y^1
- i0)], \nonumber\\ && a^2\,D^{(+)}_L(y^1 - x^1; M) +
b^2\,D^{(+)}_R(y^1 - x^1; M) =\nonumber\\ &&= -\frac{a^2}{4\pi}\,{\ell
n}[iM(y^1 - x^1 - i0)] - \frac{b^2}{4\pi}\,{\ell n}[- iM(y^1 - x^1 -
i0)] =\nonumber\\ &&= - i\,\frac{a^2 - b^2}{8} - \frac{a^2 +
b^2}{4\pi}\,{\ell n}[M(y^1 - x^1 - i0)].
\end{eqnarray}
Using relations (\ref{label5.13}) we derive the anti--commutation
relation
\begin{eqnarray}\label{label5.14}
\hspace{-0.5in}&&\{\psi_1(x^0,x^1), \psi^{\dagger}_1(x^0,y^1)\}
=\psi_1(x^0,x^1) \psi^{\dagger}_1(x^0,y^1) +
\psi^{\dagger}_1(x^0,y^1)\psi_1(x^0,x^1) =
\frac{M}{2\pi}\,e^{\textstyle - i(a^2 - b^2)/8} \nonumber\\
\hspace{-0.5in}&&\times\,\Big[(x^1 - y^1 - i0)^{\textstyle -(a^2 +
b^2)/4\pi} + (y^1 - x^1 - i0)^{\textstyle - (a^2 + b^2)/4\pi}
\,\Big]\,M^{\textstyle - (a^2 + b^2)/4\pi}\nonumber\\
\hspace{-0.5in}&&\times\,:\!e^{\textstyle +i(a\,\varphi_L(x^0, x^1) +
b\,\varphi_R(x^0, x^1)) - i(a\,\varphi_L(x^0,y^1) +
b\,\varphi_R(x^0,y^1))}\!:.
\end{eqnarray}
The r.h.s. of (\ref{label5.14}) is not equal to $\delta(x^1 - y^1)$
due to the distinction of the dynamical dimension of the massless
Thirring fermion fields from the canonical dimension $D_{\psi} =
1/2$. In order to get the canonical anti--commutation relations we
have to subtract the contribution of the dynamical dimension. For this
aim we have to introduce the products
\begin{eqnarray}\label{label5.15}
\Psi_1(x^0,x^1)\Psi^{\dagger}_1(x^0,y^1) &=& (-1)^n\,[M(x^1 - y^1 -
i0)]^{\textstyle (a^2 + b^2 - 4\pi)/4\pi}\psi_1(x^0,x^1)
\psi^{\dagger}_1(x^0,y^1),\nonumber\\
\Psi^{\dagger}_1(x^0,y^1)\Psi_1(x^0,x^1) &=& (-1)^n\,[M(y^1 - x^1 -
i0)]^{\textstyle (a^2 + b^2 -
4\pi)/4\pi}\psi^{\dagger}_1(x^0,y^1)\psi_1(x^0,x^1),\nonumber\\ &&
\end{eqnarray}
where $n\in Z$. Following [20] and setting $a^2 - b^2 = 4\pi(2n + 1)$
we get for the fermion fields $\Psi_1(x^0,x^1)$ and
$\Psi^{\dagger}_1(x^0,y^1)$ the canonical equal--time
anti--commutation relation
\begin{eqnarray}\label{label5.16}
\{\Psi_1(x^0,x^1),\Psi^{\dagger}_1(x^0,y^1)\} = \delta(x^1 - y^1).
\end{eqnarray}
The other standard anti--commutation relations are
\begin{eqnarray}\label{label5.17}
&&\{\psi_1(x^0,x^1),\psi_2(y^0,y^1)\}= 0,\nonumber\\
&&\{\psi_2(x^0,x^1),\psi_2(y^0,y^1)\}= 0,\nonumber\\
&&\{\Psi_2(x^0,x^1), \Psi^{\dagger}_2(x^0,y^1)\} = \delta(x^1 - y^1)
\end{eqnarray}
and can be derived in a similar way.

For $x^1 \sim y^1$ the Wightman functions (\ref{label5.13}) can be
rewritten as follows
\begin{eqnarray}\label{label5.18}
&& a^2\,D^{(+)}_L(x^1 - y^1; M) + b^2\,D^{(+)}_R(x^1 - y^1; M)
=\nonumber\\ &&= -\frac{a^2}{4\pi}\,{\ell n}[iM(x^1 - y^1 - i0)] -
\frac{b^2}{4\pi}\,{\ell n}[- iM(x^1 - y^1 - i0)] =\nonumber\\ &&= -
i\,\frac{a^2 - b^2}{8} - \frac{a^2 + b^2 - 4\pi}{4\pi}\,{\ell
n}\Big(\frac{M}{\Lambda}\Big) - {\ell n}[M(x^1 - y^1 - i0)],
\nonumber\\ && a^2\,D^{(+)}_L(y^1 - x^1; M) + b^2\,D^{(+)}_R(y^1 -
x^1; M) =\nonumber\\ &&= -\frac{a^2}{4\pi}\,{\ell n}[iM(y^1 - x^1 -
i0)] - \frac{b^2}{4\pi}\,{\ell n}[- iM(y^1 - x^1 - i0)] =\nonumber\\
&&= - i\,\frac{a^2 - b^2}{8} - \frac{a^2 + b^2 - 4\pi}{4\pi}\,{\ell
n}\Big(\frac{M}{\Lambda}\Big) - {\ell n}[M(y^1 - x^1 - i0)],
\end{eqnarray}
where $\Lambda$ is the ultra--violet cut--off [4]. Using
(\ref{label5.18}) the anti--commutator (\ref{label5.12}) is
represented by
\begin{eqnarray}\label{label5.19}
\hspace{-0.5in}&&\{\psi_1(x^0,x^1), \psi^{\dagger}_1(x^0,y^1)\} =
\frac{1}{2\pi}\,\Big(\frac{\Lambda}{M}\Big)^{\textstyle (a^2 + b^2
-4\pi)/4\pi}\,e^{\textstyle - i(a^2 - b^2)/8} \nonumber\\
\hspace{-0.5in}&&\times\,\Big[\frac{1}{x^1 - y^1 - i0} + \frac{1}{y^1 -
x^1 - i0}\Big]\nonumber\\
\hspace{-0.5in}&&\times\,:\!e^{\textstyle +i(a\,\varphi_L(x^0, x^1) +
b\,\varphi_R(x^0, x^1)) - i(a\,\varphi_L(x^0,y^1) +
b\,\varphi_R(x^0,y^1))}\!: =\nonumber\\
\hspace{-0.5in}&& = \Big(\frac{\Lambda}{M}\Big)^{\textstyle (a^2 + b^2
-4\pi)/4\pi}\,\delta(x^1 - y^1).
\end{eqnarray}
where we have set $a^2 - b^2 = 4\pi$ [20]. Introducing the
renormalization constant $Z_2$ of the wave functions of the massless
Thirring fermion fields [4]
\begin{eqnarray}\label{label5.20}
Z_2 = \Big(\frac{\Lambda}{M}\Big)^{\textstyle (a^2 + b^2
-4\pi)/4\pi},
\end{eqnarray}
we can transcribe the equal--time anti--commutation relation
(\ref{label5.19}) into the standard form
\begin{eqnarray}\label{label5.21}
\hspace{-0.5in}&&\{\psi_1(x^0,x^1),
\psi^{\dagger}_1(x^0,y^1)\} = Z_2\,\delta(x^1 - y^1).
\end{eqnarray}
This agrees fully with Mandelstam's relation (see Eq.(3.1) of
Ref.[17]). Recall, that according to [4] the dynamical dimension of the
massless Thirring fermion fields in the boson representation defined
by (\ref{label5.1}) is equal to
\begin{eqnarray}\label{label5.22}
d_{\psi} = \frac{a^2 + b^2}{8\pi} - \frac{1}{2}.
\end{eqnarray}
This completes the analysis of the Thirring fermion fields in the
boson field representation.

\section{Fermion condensation in the boson field representation 
of the massless Thirring fermion fields}
\setcounter{equation}{0}

\hspace{0.2in} In order to show that the Thirring fermion fields in
the boson field representation (\ref{label5.1}) are quantized in the
chirally broken phase, it is sufficient to calculate the vacuum
expectation value of the fermion operator $\bar{\psi}(x)\psi(x)$,
expressed in terms of the boson fields.

In the boson field representation (\ref{label5.1}) the operator
$\bar{\psi}(x)\psi(x)$ reads (see the Appendix)
\begin{eqnarray}\label{label6.1}
\hspace{-0.3in}&&\bar{\psi}(x)\psi(x) = \psi^{\dagger}_2(x)\psi_1(x) +
\psi^{\dagger}_1(x)\psi_2(x) = -\,\frac{M}{2\pi}\,:\!e^{\textstyle
-i(b\,\varphi_L(x) + a\,\varphi_R(x))}\!:\nonumber\\
\hspace{-0.3in}&&\times\,e^{\textstyle -i\,\pi\,((a +
b)/4ab)\,[Q_L(0) + Q_R(0)]}\,:\!e^{\textstyle
+i(a\,\varphi_L(x) + b\,\varphi_R(x))}\!:\nonumber\\
\hspace{-0.3in}&& - \,\frac{M}{2\pi}\, :\!e^{\textstyle -
i(a\,\varphi_L(x) + b\,\varphi_R(x))}\!:\,e^{\textstyle +i\,\pi\,((a +
b)/4ab)\,[Q_L(0) + Q_R(0)]}\nonumber\\
\hspace{-0.3in}&&\times\,:\!e^{\textstyle +i(b\,\varphi_L(x) +
a\,\varphi_R(x))}\!:\, = \nonumber\\
\hspace{-0.3in}&&= -\,\frac{M}{2\pi}\,e^{\textstyle +\,i\pi\,(a +
b)^2/4ab}\,e^{\textstyle +\,ab\,[D^{(+)}_L(0; M) + D^{(+)}_R(0; M)]}
\nonumber\\
\hspace{-0.3in}&&\times\,e^{\textstyle -\,i\,\pi\,((a +
b)/4ab)\,[Q_L(0) + Q_R(0)]}\,:\!e^{\textstyle -\,i\,(b -
a)\,\varphi_L(x) + \,i\,(b - a)\,\varphi_R(x)}\!:\nonumber\\
\hspace{-0.3in}&&-\,\frac{M}{2\pi}\,e^{\textstyle -\,i\pi\,(a +
b)^2/4ab}\,e^{\textstyle +\,ab\,[D^{(+)}_L(0; M) + D^{(+)}_R(0; M)]}
\nonumber\\
\hspace{-0.3in}&&\times\,:\!e^{\textstyle -\,i\,(b - a)\,\varphi_L(x)
+ \,i\,(b - a)\,\varphi_R(x)}\!:\,e^{\textstyle +\,i\,\pi\,((a +
b)/4ab)\,[Q_L(0) + Q_R(0)]}.
\end{eqnarray}
For the calculation of the vacuum expectation value of the operator
(\ref{label6.1}) we have to use the {\it fermion} vacuum
$|\Omega_T\rangle$ introduced by Morchio, Pierotti and Strocchi
[20]. Using the rules for the calculation of vacuum expectation values
with respect to the {\it fermion} vacuum $|\Omega_T\rangle$,
formulated by Morchio, Pierotti and Strocchi [20], we get 
\begin{eqnarray}\label{label6.2}
\hspace{-0.3in}\langle \Omega_T|\bar{\psi}(x)\psi(x)|\Omega_T\rangle =
-\,\frac{M}{\pi}\,\cos\Big(\pi\frac{(a +
b)^2}{4ab}\,\Big)\,e^{\textstyle +\,ab\,[D^{(+)}_L(0;M) +
D^{(+)}_R(0;M)]}.
\end{eqnarray}
The sum of the Wightman functions $D^{(+)}(0; M) = D^{(+)}_L(0; M) +
D^{(+)}_R(0; M)$ is equal to [4]
\begin{eqnarray}\label{label6.3}
D^{(+)}(0; M) = D^{(+)}_L(0; M) + D^{(+)}_R(0; M) =
\frac{1}{2\pi}\,{\ell n}\Big(\frac{\Lambda}{M}\Big),
\end{eqnarray}
where $\Lambda$ is the ultra--violet cut--off [4].

Substituting (\ref{label6.3}) in (\ref{label6.2}) we arrive at the
fermion condensate
\begin{eqnarray}\label{label6.4}
\langle \Omega_T|\bar{\psi}(x)\psi(x)|\Omega_T\rangle =
-\,\frac{M}{\pi}\,\cos\Big(\pi\frac{(a +
b)^2}{4ab}\,\Big)\,\Big(\frac{\Lambda}{M}\Big)^{\textstyle ab/2\pi}.
\end{eqnarray}
For the calculation of the vacuum expectation value of the operator
$\bar{\psi}(x)\psi(x)$ with respect to the vacuum constructed in [2,3]
it is convenient to rewrite the expression (\ref{label6.1}) as follows
\begin{eqnarray}\label{label6.5}
\hspace{-0.3in}&&\bar{\psi}(x)\psi(x) =
-\,\frac{M}{2\pi}\,e^{\textstyle +\,i\pi\,(a +
b)^2/4ab}\,e^{\textstyle +\,ab\,[D^{(+)}_L(0) + D^{(+)}_R(0)]}
\nonumber\\
\hspace{-0.3in}&&\times\,e^{\textstyle -\,i\,\pi\,((a +
b)/2ab)\,Q(0)}\,:\!e^{\textstyle +\,i\,(b -
a)\,\tilde{\varphi}(x)}\!:\nonumber\\
\hspace{-0.3in}&&-\,\frac{M}{2\pi}\,e^{\textstyle -\,i\pi\,(a +
b)^2/4ab}\,e^{\textstyle +\,ab\,[D^{(+)}_L(0; M) + D^{(+)}_R(0; M)]}
\nonumber\\
\hspace{-0.3in}&&\times\,:\!e^{\textstyle +\,i\,(b -
a)\,\tilde{\varphi}(x)}\!:\,e^{\textstyle +\,i\,\pi\,((a +
b)/2ab)\,Q(0)}.
\end{eqnarray}
The vacuum expectation value of the operator $\bar{\psi}(x)\psi(x)$ is
defined by
\begin{eqnarray}\label{label6.6}
\hspace{-0.3in}&&\langle 0|\bar{\psi}(x)\psi(x)|0\rangle =
-\,\frac{M}{2\pi}\,e^{\textstyle +\,i\pi\,(a +
b)^2/4ab}\,e^{\textstyle +\,ab\,[D^{(+)}_L(0) + D^{(+)}_R(0)]}
\nonumber\\
\hspace{-0.3in}&&\times\,\langle 0|e^{\textstyle -\,i\,\pi\,((a +
b)/2ab)\,Q(0)}|0\rangle\,\langle 0|:\!e^{\textstyle +\,i\,(b -
a)\,\tilde{\varphi}(x)}\!:|0\rangle\nonumber\\
\hspace{-0.3in}&&-\,\frac{M}{2\pi}\,e^{\textstyle -\,i\pi\,(a +
b)^2/4ab}\,e^{\textstyle +\,ab\,[D^{(+)}_L(0; M) + D^{(+)}_R(0; M)]}
\nonumber\\
\hspace{-0.3in}&&\times\,\langle 0|:\!e^{\textstyle +\,i\,(b -
a)\,\tilde{\varphi}(x)}\!:|0\rangle\,\langle 0|e^{\textstyle
+\,i\,\pi\,((a + b)/2ab)\,Q(0)}|0\rangle.
\end{eqnarray}
Using the properties of the vacuum wave function $|0\rangle$
investigated in [2,3] we get
\begin{eqnarray}\label{label6.7}
\hspace{-0.3in}&&\langle 0|\bar{\psi}(x)\psi(x)|0\rangle =
-\,\frac{M}{\pi}\,\cos\Big(\pi\,\frac{(a +
b)^2}{4ab}\Big)\,\Big(\frac{\Lambda}{M}\Big)^{\textstyle
ab/2\pi}\,e^{\textstyle -\pi^2\,(a + b)^2/16a^2b^2},
\end{eqnarray}
where we have applied (\ref{label6.3}).

The obtained results (\ref{label6.4}) and (\ref{label6.7}) testify the
equivalence of the solution of the massless Thirring model with
fermion fields constructed from the free massless boson fields to the
massless Thirring model with fermion fields quantized in the chirally
broken phase.

\section{Constant of motion of the massless Thirring model in 
the boson field representation}
\setcounter{equation}{0}

\hspace{0.2in} In the boson field representation the fermion operator
products defining the constant of motion of the massless Thirring
model read (see the Appendix)
\begin{eqnarray}\label{label7.1}
&&\psi^{\dagger}_1(x)\psi_2(x)\psi^{\dagger}_2(x)\psi_1(x) +
\psi^{\dagger}_2(x)\psi_1(x)\psi^{\dagger}_1(x)\psi_2(x) =\nonumber\\
&&= \frac{M^2}{4\pi^2}\,:\!e^{\textstyle -\,i\,(a\,\varphi_L(x) +
b\,\varphi_R(x))}\!:\,e^{\textstyle +i\,(\pi/4)\,(b^{-1}Q_L(0) +
a^{-1}Q_R(0))}\nonumber\\ &&\times\,e^{\textstyle
+i\,(\pi/4)\,(a^{-1}Q_L(0) + b^{-1}Q_R(0))}\,:\!e^{\textstyle
+\,i\,(b\,\varphi_L(x) + a\,\varphi_R(x))}\!:\nonumber\\
&&\times\,:\!e^{\textstyle -\,i\,(b\,\varphi_L(x) +
a\,\varphi_R(x))}\!:\,\,e^{\textstyle -i\,(\pi/4)\,(a^{-1}Q_L(0) +
b^{-1}Q_R(0))}\nonumber\\ &&\times\,e^{\textstyle
-i\,(\pi/4)\,(b^{-1}Q_L(0) + a^{-1}Q_R(0))}\,:\!e^{\textstyle
+\,i\,(a\,\varphi_L(x) + b\,\varphi_R(x))}\!:\nonumber\\ 
&&+  \frac{M^2}{4\pi^2}\,:\!e^{\textstyle -\,i\,(b\,\varphi_L(x) +
a\,\varphi_R(x))}\!:\,e^{\textstyle -i\,(\pi/4)\,(a^{-1}Q_L(0) +
b^{-1}Q_R(0))}\nonumber\\ &&\times\,e^{\textstyle
-i\,(\pi/4)\,(b^{-1}Q_L(0) + a^{-1}Q_R(0))}\,:\!e^{\textstyle
+\,i\,(a\,\varphi_L(x) + b\,\varphi_R(x))}\!:\nonumber\\
&&\times\,:\!e^{\textstyle -\,i\,(a\,\varphi_L(x) +
b\,\varphi_R(x))}\!:\,\,e^{\textstyle +i\,(\pi/4)\,(b^{-1}Q_L(0) +
a^{-1}Q_R(0))}\nonumber\\ &&\times\,e^{\textstyle
+i\,(\pi/4)\,(a^{-1}Q_L(0) + b^{-1}Q_R(0))}\,:\!e^{\textstyle
+\,i\,(b\,\varphi_L(x) + a\,\varphi_R(x))}\!:.
\end{eqnarray}
One can easily show that the r.h.s. of (\ref{label7.1}) is equal to
\begin{eqnarray}\label{label7.2}
\psi^{\dagger}_1(x)\psi_2(x)\psi^{\dagger}_2(x)\psi_1(x) +
\psi^{\dagger}_2(x)\psi_1(x)\psi^{\dagger}_1(x)\psi_2(x) =
\frac{M^2}{2\pi^2}\,e^{\textstyle \,(a^2 + b^2)\,D^{(+)}(0; M)}.
\end{eqnarray}
Using the definition of the Wightman function $D^{(+)}(0; M)$ given by
(\ref{label6.3}) we get
\begin{eqnarray}\label{label7.3}
\psi^{\dagger}_1(x)\psi_2(x)\psi^{\dagger}_2(x)\psi_1(x) +
\psi^{\dagger}_2(x)\psi_1(x)\psi^{\dagger}_1(x)\psi_2(x) =
\frac{M^2}{2\pi^2}\,\Big(\frac{\Lambda}{M}\Big)^{\textstyle (a^2 +
b^2)/2\pi}.
\end{eqnarray}
This defines the constant $C$ in terms of the normalization scale $M$
and the ultra--violet cut--off $\Lambda$
\begin{eqnarray}\label{label7.4}
C = \frac{M^2}{\pi^2}\,\Big(\frac{\Lambda}{M}\Big)^{\textstyle (a^2 +
b^2)/2\pi}.
\end{eqnarray}
Thus, we have shown that the boson field representation, developed by
Morchio, Pierotti and Strocchi [20] and supplemented by our recipe
[2,3] for the removal of the infrared divergences in the quantum field
theory of the free massless (pseudo)scalar fields defined in
1+1--dimensional space--time, confirms completely our assertion
concerning the existence of the constant of motion for the evolution
of the fermion fields described by the massless Thirring model [1].

\section{Conclusion}
\setcounter{equation}{0}

\hspace{0.2in} We have analysed the boson field representation of the
massless fermion fields defined in 1+1--dimensional space--time. We
have shown that the formulation of the quantum field theory of the
free massless (pseudo)scalar field, free of infrared divergences in
1+1--dimensional space--time, has given the basis for the
self--consistent description of the fermion field operators in terms
of the free massless (pseudo)scalar fields.  In our version of the
approach developed by Morchio, Pierotti and Strocchi [20] the fermion
field operators in the boson field representation do not depend on the
infrared cut--off $\mu$, which should be taken finally in the limit
$\mu \to 0$. We would like to emphasize that such a limit would make
all the results ill--defined.  The most important disagreement, which
could be produced by the limit $\mu \to 0$, would be the violation of
the constant of motion revealed in [1]. As has been shown in [1] the
product of the massless Thirring fermion field operators
$[\bar{\psi}(x)\psi(x)]^2 + [\bar{\psi}(x)i\gamma^5\psi(x)]^2 = C$ is
conserved in the evolution of the fermion fields obeying the equations
of motion for the massless Thirring model. The constant $C$ should not
vanish in the limit $\mu \to 0$.

We have shown that the Thirring fermion field operators defined in the
boson field representation by the boson fields, described by a quantum
field theory free of infrared divergences [2,3], satisfy the constant
of motion with the constant $C$ independent of the infrared cut--off.

The non--vanishing value of the fermion condensate obtained within the
boson field representation agrees fully with our assertions concerning
(i) the existence of the chirally broken phase in the massless
Thirring model [1,4] and (ii) the equivalence of the fermion fields,
represented by the massless boson fields, to the Thirring fermion
fields quantized in the chirally broken phase.

\section*{Acknowledgement}

\hspace{0.2in} We are grateful to Ludmil Hadjiivanov for helpful
discussions and for reading the manuscript.

\section*{Appendix. Normal ordering of exponential operators 
depending on free massless boson fields}

\hspace{0.2in} The normal ordering of the exponential operators
depending on the free massless boson fields defined in the quantum
field theory without infrared divergences, where the zero--mode
configurations are removed by the constraint on the external sources
$$
\int d^2x\,J(x) = \tilde{J}(0) = 0, \eqno({\rm A}.1)
$$
can not be understood and performed naively. In the case of the naive
performance of the normal ordering of the exponential operators in the
massless boson field representation one should encounter the problem
of the appearance of the infrared cut--off $\mu$ in the final
expressions. This should contradict the results of the calculation of
the vacuum expectation values of these operators. Therefore, the
required normal ordering of the exponential operators in the massless
boson field representation should be carried out in the way agreeing
with the calculation of vacuum expectation values, which do not depend
on the infrared cut--off $\mu$ [2].

Let us describe the required procedure of the normal ordering of the
exponential operators. We suggest to start with the operator ${\cal
O}^{(1)}[\varphi]$ defined by
$$
{\cal O}^{(1)}[\varphi] = e^{\textstyle
i\beta\,\varphi(x)}.\eqno({\rm A}.2)
$$
The direct application of Wick's theorem [21]\,\footnote{see also
Glimm and Jaffe, p.107 of Ref.[11] and Eq.(2.6) of Ref.[22].} allows
to represent the operator ({\rm A}.2) in the following normal--ordered
form
$$
e^{\textstyle i\beta\,\varphi(x)} = :e^{\textstyle
i\beta\,\varphi(x)}:\,e^{\textstyle \frac{1}{2}\,\beta^2\,i\Delta(0)},
\eqno({\rm A}.3)
$$
where the symbol $:\ldots:$ stands for the normal ordering and the
Green function $i\Delta(0)$ will be defined later.

The vacuum expectation value of the both sides of the relation
({\rm A}.3) is equal to
$$
\langle 0|e^{\textstyle i\beta\,\varphi(x)}|0\rangle = \langle
0|:e^{\textstyle i\beta\,\varphi(x)}:|0\rangle\,e^{\textstyle
\frac{1}{2}\,\beta^2\,i\Delta(0)},\eqno({\rm A}.4)
$$
Since the vacuum expectation value of the normal--ordered exponential
is a unity, the relation ({\rm A}.4) reads
$$
\langle 0|e^{\textstyle i\beta\,\varphi(x)}|0\rangle = e^{\textstyle
\frac{1}{2}\,\beta^2\,i\Delta(0)},\eqno({\rm A}.5)
$$
The l.h.s. of this relation can be calculated in terms of the
generating functional of Green functions $Z[J]$ (\ref{label1.12})
which does not depend on the infrared cut--off $\mu$. We get [2]
$$
\langle 0|e^{\textstyle i\beta\,\varphi(x)}|0\rangle = \langle
0|e^{\textstyle i\,\beta\,\varphi(0)}|0\rangle =
\exp\Big\{\beta\,\frac{\delta}{\delta J(0)}\Big\}Z[J]\Big|_{\textstyle
J=0} = e^{\textstyle \frac{1}{2}\,\beta^2\,i\Delta(0;M)}.\eqno({\rm
A}.6)
$$
The Green function $i\Delta(0;M)$ is defined by [2]
$$
i\Delta(0;M) = - \frac{1}{4\pi}\,{\ell
n}\Big(\frac{\Lambda^2}{M^2}\Big).\eqno({\rm
A}.7)
$$
Substituting ({\rm A}.6) in ({\rm A}.5) we obtain that
$$
i\Delta(0) = i\Delta(0;M) = - \frac{1}{4\pi}\,{\ell
n}\Big(\frac{\Lambda^2}{M^2}\Big).\eqno({\rm A}.8)
$$
Hence, the normal--ordered form of the operator $e^{\textstyle
i\beta\,\varphi(x)}$ is defined by
$$
e^{\textstyle i\beta\,\varphi(x)} = :e^{\textstyle
i\beta\,\varphi(x)}:\,e^{\textstyle \frac{1}{2}\,\beta^2\,i\Delta(0;M)}.
\eqno({\rm A}.9)
$$
The inverse expression reads 
$$
:e^{\textstyle i\beta\,\varphi(x)}: = e^{\textstyle -
\frac{1}{2}\,\beta^2\,i\Delta(0;M)}\,e^{\textstyle
i\beta\,\varphi(x)}.\eqno({\rm A}.10)
$$
Now let us consider the operator ${\cal O}^{(2)}[\varphi]$ given by
$$
{\cal O}^{(2)}[\varphi] = e^{\textstyle
i\alpha\,\varphi(x)}\,e^{\textstyle i\beta\,\varphi(y)}.  \eqno({\rm
A}.11)
$$
Using ({\rm A}.9) we get 
$$
e^{\textstyle i\alpha\,\varphi(x)}\,e^{\textstyle i\beta\,\varphi(y)}
= :e^{\textstyle i\alpha\,\varphi(x)}::e^{\textstyle
i\beta\,\varphi(y)}:e^{\textstyle +\frac{1}{2}\,(\alpha^2 +
\beta^2)\,i\Delta(0;M)}.\eqno({\rm A}.12)
$$
The product of the normal--ordered exponentials is equal to
$$
:e^{\textstyle i\alpha\,\varphi(x)}::e^{\textstyle
i\beta\,\varphi(y)}: = e^{\textstyle - \frac{1}{2}\,(\alpha^2 +
\beta^2)\,i\Delta(0;M)}\,e^{\textstyle
i\alpha\,\varphi(x)}\,e^{\textstyle i\beta\,\varphi(y)}.\eqno({\rm
A}.13)
$$
Taking time--ordered expressions and calculating the vacuum
expectation values of the both sides we get
$$
\langle 0|{\rm T}\Big(:e^{\textstyle
i\alpha\,\varphi(x)}::e^{\textstyle i\beta\,\varphi(y)}:\Big)|0\rangle
= e^{\textstyle -\frac{1}{2}\,(\alpha^2 +
\beta^2)\,i\Delta(0;M)}\,\langle 0|{\rm T}\Big(e^{\textstyle
i\alpha\,\varphi(x)}\,e^{\textstyle i\beta\,\varphi(y)}\Big)|0\rangle
.\eqno({\rm A}.14)
$$
The vacuum expectation value of the time--ordered operator in the
r.h.s. of ({\rm A}.14) can be calculated in terms of $Z[J]$. The
result reads [2]
$$
\langle 0|{\rm T}\Big(e^{\textstyle
i\alpha\,\varphi(x)}\,e^{\textstyle i\beta\,\varphi(y)}\Big)|0\rangle
= \exp\Big\{-i\alpha \frac{\delta }{\delta J(x)} - i\beta \frac{\delta
}{\delta J(y)}\Big\}Z[J]\Big|_{\textstyle J = 0}=
$$
$$
= e^{\textstyle +\frac{1}{2}\,(\alpha^2 + \beta^2)\,i\Delta(0;
M)}\,e^{\textstyle +\alpha\,\beta\,i\Delta(x - y; M)}.\eqno({\rm
A}.15)
$$
Substituting ({\rm A}.15) in ({\rm A}.14) we get
$$
\langle 0|{\rm T}\Big(:e^{\textstyle
i\alpha\,\varphi(x)}::e^{\textstyle i\beta\,\varphi(y)}:\Big)|0\rangle
= e^{\textstyle +\alpha\,\beta\,i\Delta(x - y; M)}.\eqno({\rm A}.16)
$$
This yields
$$
\langle 0|:e^{\textstyle i\alpha\,\varphi(x)}::e^{\textstyle
i\beta\,\varphi(y)}:|0\rangle = e^{\textstyle
-\alpha\,\beta\,D^{(+)}(x - y; M)}.\eqno({\rm A}.17)
$$
Hence, taking into account that the vacuum expectation value of the
normal--ordered exponential is equal to unity, the operator form of
the relation ({\rm A}.17) reads
$$
:e^{\textstyle i\alpha\,\varphi(x)}::e^{\textstyle
i\beta\,\varphi(y)}: = e^{\textstyle -\alpha\,\beta\,D^{(+)}(x - y;
M)}\,:e^{\textstyle i\alpha\,\varphi(x)
+i\beta\,\varphi(y)}:.\eqno({\rm A}.18)
$$
This relation can be rewritten in the form
$$
:e^{\textstyle i\alpha\,\varphi(x)}::e^{\textstyle
i\beta\,\varphi(y)}: = e^{\textstyle
-\alpha\,\beta\,[\varphi^{(+)}(x),\varphi^{(-)}(y)]}\,:e^{\textstyle
i\alpha\,\varphi(x) +i\beta\,\varphi(y)}:,\eqno({\rm A}.19)
$$
where the commutator $[\varphi^{(+)}(x),\varphi^{(-)}(y)]$ is defined
by
$$
[\varphi^{(+)}(x),\varphi^{(-)}(y)] = D^{(+)}(x - y; M).\eqno({\rm
A}.19)
$$
Thus, we have formulated the procedure for the normal ordering
exponential operators in the boson field representation of fermion
fields with the boson fields described by the quantum field theory
without infrared divergences [2]. Following this procedure one can
calculate any product of the normal--ordered exponential operators
depending on the free massless boson fields described by the quantum
field theory without infrared divergences. 

We would like to emphasize that according to the normal ordering
procedure, formulated above, no infrared cut--off $\mu$ can appear in
the products of fermion field operators. All expressions should depend
only on the finite arbitrary scale $M$.

\end{document}